\documentclass{ptephy}

\usepackage{graphicx}

\begin{document}

\title{IceCube potential for detecting $Q$-ball dark matter in gauge mediation}

\author{\name{\fname{Shinta} \surname{Kasuya}}{1 \, \ast}, \name{\fname{Masahiro} \surname{Kawasaki}}{2,3}, 
and \name{\fname{Tsutomu} \midname{T.} \surname{Yanagida}}{3}}

\address{\affil{1}{Department of Mathematics and Physics,
     Kanagawa University, Kanagawa 259-1293, Japan}
\affil{2}{Institute for Cosmic Ray Research, the University of Tokyo, Chiba 277-8582, Japan}
\affil{3}{Kavli Institute for the Physics and Mathematics of the Universe (WPI), 
  Todai Institutes for Advanced Study, the University of Tokyo, Chiba 277-8582, Japan}
\email{kasuya@kanagawa-u.ac.jp}}

\begin{abstract}%
We study $Q$-ball dark matter in gauge-mediated supersymmetry breaking, and seek the
possibility of detection in the IceCube experiment. We find that the $Q$ balls would be the dark matter in 
the parameter region different from that for gravitino dark matter. In particular, the $Q$ ball is a good 
dark matter candidate for low reheating temperature, which may be suitable for the Affleck-Dine 
baryogenesis and/or nonthermal leptogenesis. Dark matter $Q$ balls are detectable by IceCube-like 
experiments in the future, which is the peculiar feature compared to the case of gravitino dark matter.
\end{abstract}

\subjectindex{}

\maketitle

\tableofcontents

\section{Introduction}
Gauge mediation \cite{GMSB1,GMSB2,GMSB3,GMSB4,GMSB5,GMSB6,GMSB7} is very attractive
for the supersymmetry (SUSY) breaking mechanism, 
since it is free from the serious flavor-changing neutral current (FCNC) problem 
in the supersymmetric standard model. It has been recently pointed out that the minimal gauge 
mediation model is still consistent with all experimental data and cosmological constraints \cite{HIYY}.
Thus it is worthwhile to discuss possible dark matter candidates in the gauge mediation model. 

A well-known candidate is the gravitino. 
If the gravitino mass $m_{3/2}$ is lighter than 1 keV, the gravitino cannot saturate the dark 
matter abundance \cite{MMY}. Even for 1 keV $\lesssim m_{3/2} \lesssim$ 100 keV, the gravitino dark 
matter would be too warm, and the small-scale fluctuations will be erased \cite{KFY,Viel}. In addition, 
the reheating temperature after inflation should be $T_{\rm RH} \gtrsim 10^3$~GeV for the thermally 
produced gravitinos to be dark matter \cite{KKMY}.

To test the gravitino dark matter we need first to produce 
SUSY particles in collider experiments and see their decays into the almost massless gravitino. 
However, even if the SUSY particles are produced it is very challenging to observe the decay if the 
gravitino is heavier than 1 keV.

In this paper we discuss the other possibility that the $Q$ ball is the dark matter 
in gauge mediation, and look for the discovery potential of $Q$-ball dark matter at IceCube.
There are a lot of flat directions in SUSY. They consist of some combinations of
squarks (and sleptons) in the minimal supersymmetric standard model (MSSM), so that they carry 
baryonic charge. The $Q$ ball is a nontopological soliton, the energy minimum configuration of the flat 
direction for a finite baryon number \cite{Coleman}. Since the $Q$ ball with large enough charge 
(the baryon number) is stable against decay into nucleons in gauge mediation, 
it is reasonable to consider the $Q$ ball as the dark matter candidate \cite{KuSh}.

We show that the $Q$ ball would be the dark matter in the parameter region different
from that for gravitino dark matter: the gravitino mass could range for keV $\lesssim m_{3/2} \lesssim$ GeV,
or even smaller, and the reheating temperature must be $T_{\rm RH} \lesssim 10^4$~GeV.
Moreover, it has a completely different detection procedure from other dark matter candidates. 
Very large volume neutrino detectors such as IceCube can directly detect dark matter $Q$ balls 
in the (near) future.

The sketch of the paper is as follows. In the next section, we show the set-up of gauge-mediated 
SUSY breaking. In Sect.~3, we review the $Q$ ball in gauge mediation, where there are two types
of $Q$ ball. We estimate the abundance of the $Q$ ball and show the parameter region where
the $Q$ balls could be the dark matter in Sect.~4. In the same section, we look for the possibility 
of $Q$-ball detection in the IceCube experiment. Sect.~5 is devoted to our conclusion.

\section{Gauge-mediated SUSY breaking}
We adopt the so-called minimal direct gauge mediation described only by a few parameters, and
free from both the SUSY FCNC problem and the SUSY CP problem \cite{HIYY}.
The SUSY is spontaneously broken by the vacuum expectation value of the SUSY-breaking field $Z$ as
\begin{equation}
\langle Z \rangle =0, \qquad \langle F_Z \rangle = F.
\end{equation}
The SUSY-breaking sector is connected to the observable sector by the messenger fields 
$\Psi$ and $\bar{\Psi}$, a pair of some representations and anti-representations of the minimal
GUT group $SU(5)$, through the Yukawa interactions as
\begin{equation}
W=kZ\bar{\Psi}\Psi+M_{\rm mess}\bar{\Psi}\Psi,
\end{equation}
where $k$ is a coupling constant, and $M_{\rm mess}$ is the messenger mass. Then the MSSM gaugino
and scalar masses are estimated as
\begin{equation}
M_{\rm gaugino} \sim M_{\rm scalar} \sim \frac{g^2}{16\pi^2} \frac{kF}{M_{\rm mess}}.
\end{equation}
On the other hand, the gravitino mass is given by
\begin{equation}
m_{3/2} = \frac{F}{\sqrt{3}M_{\rm P}},
\end{equation}
where $M_{\rm P}\simeq 2.4\times 10^{18}$~GeV is the Planck mass. 
The Higgs boson mass at around 126~GeV leads to \cite{HIYY,mGMSB}
\begin{equation}
\Lambda_{\rm mess} \equiv \frac{kF}{M_{\rm mess}} \gtrsim 5 \times 10^5 \ {\rm GeV}.
\end{equation}
Since $kF<M^2$, it results in
\begin{equation}
\sqrt{k F} > \Lambda_{\rm mess} \gtrsim 5 \times 10^{5} \ {\rm GeV}.
\label{Lambdamess}
\end{equation}

\section{$Q$ balls in gauge mediation}
A $Q$ ball is a nontopological soliton, the energy-minimum configuration of the scalar field for
finite charge $Q$ \cite{Coleman}. In MSSM, this scalar field corresponds to some flat direction, which consists of
some combination of squarks (and sleptons), and the charge $Q$ would be the baryon number.
In gauge mediation, a $Q$ ball with large enough charge is stable against decay into
nucleons; as will be seen later, it can be the dark matter of the universe. 
The potential of the flat direction is expressed as \cite{KuSh,flat,EnMc}
\begin{equation}
V = V_{\rm gauge} + V_{\rm grav} = M_F^4 \left(\log\frac{|\Phi|^2}{M_{\rm mess}^2}\right)^2 
+m_{3/2}^2\left(1+K\log\frac{|\Phi|^2}{M_*^2}\right) |\Phi|^2.
\label{pot}
\end{equation}
Here the first term comes from gauge mediation effects, and $M_F$ is related to SUSY-breaking 
$F$-term as \cite{flat}
\begin{equation}
M_F \simeq \frac{g^{1/2}}{4\pi} \sqrt{kF},
\end{equation}
where $g$ generically denotes the gauge coupling of the standard model. Thus the gravitino mass
can be expressed as
\begin{equation}
m_{3/2} = \frac{1}{\sqrt{3}M_{\rm P}}\frac{(4\pi)^2 M_F^2}{gk}.
\label{m32MF}
\end{equation}
The second term of Eq.(\ref{pot}) originates from gravity mediation effects, and the one-loop correction is
included. $K$ is typically negative, $|K|=0.01-0.1$, and $M_*$ is the renormalization scale. 

Since the messenger scale $\Lambda_{\rm mess}$ has the lower limit in Eq.(\ref{Lambdamess}), it should be 
\begin{equation}
M_F \gtrsim 4 \times 10^4 g^{1/2} \, {\rm GeV}.
\label{Llimit1}
\end{equation}
Likewise, the gravitino mass has the lower bound 
\begin{equation}
m_{3/2} \gtrsim 6.1\times 10^{-8} k^{-1} \, {\rm GeV}.
\label{Llimit2}
\end{equation}
We shall call them the $\Lambda_{\rm mess}$-limit in the following.

The flat direction has large amplitude during and after inflation, and starts its oscillation 
when the Hubble parameter becomes as large as the curvature of the potential \cite{AD}. 
Once the oscillation begins, the flat direction feels spatial instabilities, which grow into $Q$ balls 
very fast \cite{KuSh,EnMc,KK1,KK2,KK3}. Actually, the field rotates in the potential due to the so-called 
$A$-terms, so the orbit of the field may be circular or oblate depending on the size of the $A$-terms. 
However, in any case, $Q$ balls (and anti-$Q$ balls in the case with the oblate orbit) form with
similar sizes \cite{KK2,KK3}, so that the $Q$-ball abundance has almost no dependence on 
the size of the $A$-terms.

There are two types of $Q$ ball: 
the gauge-meditation type and the new type \cite{new, KK3}. The former forms when the field begins 
the oscillation when the potential is dominated by $V_{\rm gauge}$, where the field amplitude at that time
is smaller than 
\begin{equation}
\phi_{\rm eq} \simeq \frac{\sqrt{2}M_F^2}{m_{3/2}},
\label{phieq}
\end{equation}
where $\displaystyle{\Phi=\frac{1}{\sqrt{2}}\phi e^{i\theta}}$. The charge of this type of $Q$ ball is 
given by \cite{KK3}
\begin{equation}
Q_{\rm G} = \beta_{\rm G} \left(\frac{\phi_{\rm osc}}{M_F}\right)^4,
\label{Qg}
\end{equation}
where $\phi_{\rm osc}$ is the amplitude of the field at the beginning of its oscillation, and 
$\beta_{\rm G}=6\times 10^{-4}$ for a circular orbit ($\varepsilon=1$), while 
$\beta_{\rm G}=6\times 10^{-5}$ for an oblate orbit ($\varepsilon\lesssim 0.1$), where $\varepsilon$
denotes the ellipticity of the field orbit. The features of the gauge-mediation type $Q$ ball are as
follows:
\begin{eqnarray}
\label{MQg}
&& M_Q \simeq \frac{4\sqrt{2}\pi}{3} \zeta M_F Q_{\rm G}^{3/4}, \\
&& R_Q \simeq \frac{1}{\sqrt{2}} \zeta^{-1} M_F^{-1} Q_{\rm G}^{1/4}, \\
\label{omegaQg}
&& \omega_Q \simeq \sqrt{2} \pi \zeta M_F Q_{\rm G}^{-1/4}, 
\end{eqnarray}
where $M_Q$ and $R_Q$ are the mass and  the size of the $Q$ ball, respectively, $\omega_Q$ is 
the rotation speed of the field inside the $Q$ ball, and $\zeta$ is the $O(1)$ parameter \cite{HNO, KY}.

On the other hand, the new type $Q$ ball is produced if $V_{\rm grav}$ dominates the potential 
when $\phi_{\rm osc} > \phi_{\rm eq}$. The charge of the $Q$ ball is estimated as \cite{new, KK3}
\begin{equation}
\label{Qn}
Q_{\rm N} = \beta_{\rm N} \left(\frac{\phi_{\rm osc}}{m_{3/2}}\right)^2,
\end{equation}
where $\beta_{\rm N} =0.02$ \cite{Hiramatsu}. The properties of this type of $Q$ ball are
\begin{eqnarray}
\label{MQn}
&& M_Q \simeq m_{3/2} Q_{\rm N}, \\
&& R_Q \simeq |K|^{-1/2} m_{3/2}^{-1}, \\
\label{omegaQn}
&& \omega_Q \simeq m_{3/2}.
\end{eqnarray}

The charge $Q$ is actually the $\Phi$-number and related to the baryon number as $B=bQ$, 
where the flat direction carries the charge $b$. For example, $b=1/3$ for the $udd$ direction.

$Q$ balls can become the dark matter of the universe if they are stable against decay into 
nucleons.\footnote{
$Q$ balls with $Q$ being the lepton number, so-called $L$ balls, can decay into neutrinos.
Large $L$ balls could have a lifetime longer than the present age of the universe, but the abundance 
becomes many orders of magnitude larger than the critical density. Therefore, $L$ balls cannot be 
the dark matter of the universe.}
$\omega_Q$ can be regarded as the effective mass of the field inside the $Q$ ball. Because of kinematics,
the $Q$ ball cannot decay into nucleons for $\omega_Q < b m_N$ with $m_N$ being the nucleon mass.
For the gauge-mediation type $Q$ ball, the condition is rewritten as
\begin{equation}
Q_{\rm G} > Q_{\rm D} \equiv 4\pi^4\zeta^4\left(\frac{M_F}{b m_N}\right)^4
\simeq 1.2 \times 10^{30} \left(\frac{\zeta}{2.5}\right)^4 \left(\frac{b}{1/3}\right)^{-4}
\left(\frac{M_F}{10^6\,{\rm GeV}}\right)^4.
\label{stableG}
\end{equation}
On the other hand, the new-type $Q$ ball is generically stable since $m_{3/2} \lesssim$ GeV.

\section{$Q$ balls as dark matter}
In this section, we investigate how the dark matter $Q$ ball can be realized, and seek the possibility 
for their direct detection. We also show the parameter region where the gravitino can be the dark matter,
and see that it resides in a different part of the parameter space.

\subsection{Dark matter $Q$ balls}
There are several conditions for the $Q$ ball to be the dark matter of the universe. 
The $Q$ ball should be (a) stable to be the dark matter, and (b) have the 
correct dark matter abundance. Since there are two types of $Q$ ball, the amplitude at the oscillation, 
$\phi_{\rm osc}$, should be in the right place of the potential, which corresponds to the condition
(c)  $\phi_{\rm osc} < \phi_{\rm eq}$ for the gauge-mediation type, and  $\phi_{\rm osc} > \phi_{\rm eq}$
for the new type $Q$ ball, so that the potential must be dominated respectively by $V_{\rm gauge}$ 
or $V_{\rm grav}$ at the onset of the field oscillation. In addition, the $\Lambda_{\rm mess}$-limit of 
(\ref{Llimit1}) or (\ref{Llimit2}) should be satisfied.

Let us first consider the gauge-mediation type $Q$ ball. 
The $Q$-ball abundance is estimated as
\begin{equation}
\frac{\rho_Q}{s} = \frac{3T_{\rm RH}}{4} \left.\frac{\rho_Q}{\rho_{\rm inf}}\right|_{\rm osc}
= \frac{3T_{\rm RH}}{4} \frac{M_Q n_\phi/Q}{3 H_{\rm osc}^2 M_{\rm P}^2} 
= \frac{3}{2} \pi\zeta \beta_{\rm G}^{-1/4} T_{\rm RH} \frac{\phi_{\rm osc}^2}{M_{\rm P}^2},
\end{equation}
where $\rho_{\rm inf}$ is the energy density of the inflaton, and $n_\phi =m_{\rm eff} \phi_{\rm osc}^2$, 
$m_{\rm eff} = 2\sqrt{2}M_F^2/\phi_{\rm osc}$, 
$3H_{\rm osc}=m_{\rm eff}$, and Eq.(\ref{MQg}) are used. Observationally, the amount of dark matter 
is $\rho_{\rm DM}/s\simeq 4.4\times 10^{-10}$~GeV \cite{Planck}. Therefore,
the amplitude of the field at the beginning of the oscillation becomes
\begin{equation}
\phi_{\rm osc} = 5.80\times 10^{12} \, {\rm GeV} 
\left(\frac{\zeta}{2.5}\right)^{-1/2} \left(\frac{\beta_{\rm G}}{6\times 10^{-4}}\right)^{1/8} 
\left(\frac{T_{\rm RH}}{\rm GeV}\right)^{-1/2}.
\label{phioscDMg}
\end{equation}
The stability condition (\ref{stableG}) and Eq.(\ref{Qg}) lead to
\begin{equation}
\phi_{\rm osc} > 2.13\times 10^{14} \, {\rm GeV} \left(\frac{\zeta}{2.5}\right)
\left(\frac{\beta_{\rm G}}{6\times 10^{-4}}\right)^{-1/4} \left(\frac{b}{1/3}\right)^{-1} 
\left(\frac{M_F}{10^6\, {\rm GeV}}\right)^2.
\end{equation}
From these two equations, we have the upper limit on the reheating temperature as
\begin{equation}
T_{\rm RH} < 7.42 \times 10^{-4} \, {\rm GeV}  \left(\frac{\zeta}{2.5}\right)^{-3}
\left(\frac{\beta_{\rm G}}{6\times 10^{-4}}\right)^{3/4} \left(\frac{b}{1/3}\right)^2 
\left(\frac{M_F}{10^6\, {\rm GeV}}\right)^{-4}.
\label{stablelimitG}
\end{equation}
for the $Q$ balls to be the dark matter.
The condition (c) can be rephrased, using Eqs.(\ref{phioscDMg}) and (\ref{phieq}) together with
Eq.(\ref{m32MF}), as the lower bound on the reheating temperature:
\begin{equation}
T_{\rm RH} > 2.44 \times 10^{-8} \, {\rm GeV} g^{-2} k^{-2} \left(\frac{\zeta}{2.5}\right)^{-1}
\left(\frac{\beta_{\rm G}}{6\times 10^{-4}}\right)^{1/4}. 
\label{phieqboundg}
\end{equation}
We show the allowed region for the $Q$-ball dark matter in Fig.1. The red line denotes the upper limit
(\ref{stablelimitG}), the magenta line shows the $\Lambda_{\rm mess}$-limit (\ref{Llimit1}) with $g=1$, 
and the blue lines represent the lower bound (\ref{phieqboundg}) for each value of $k$ shown in the 
figure. We conservatively assume that $T_{\rm RH} > 1$~MeV, so that we do not show 
$T_{\rm RH}$ lower than 1~MeV in the figures.

\begin{figure}[ht!]
\hspace{-10mm}
\includegraphics[width=90mm,angle=90]{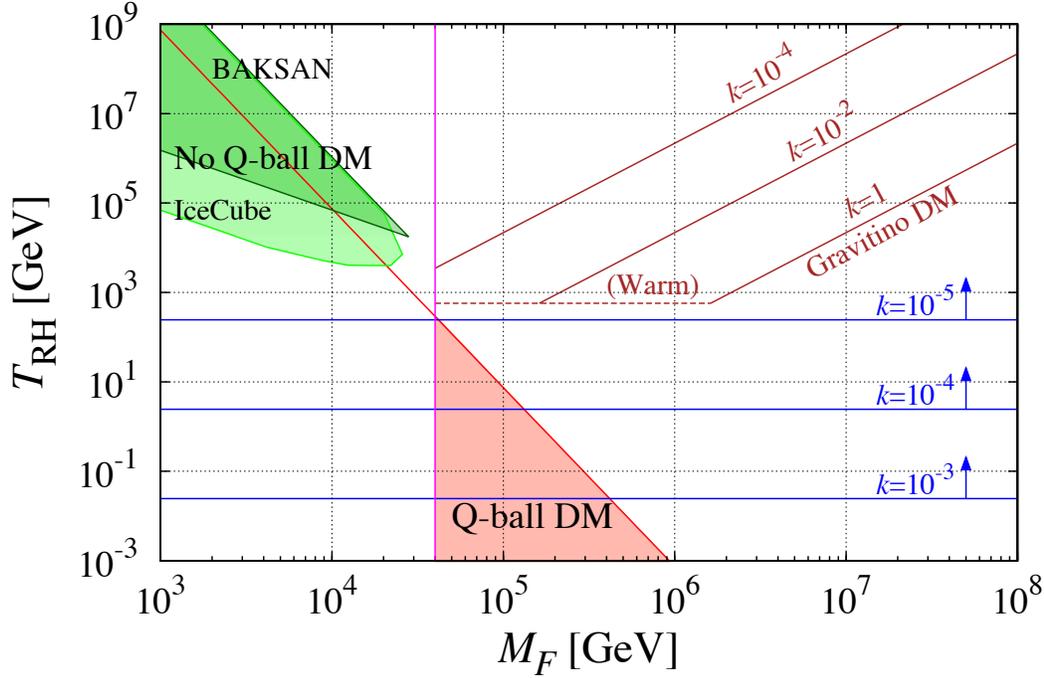} 
\vspace{10mm}
\caption{Allowed region for the gauge-mediation type $Q$ ball as the dark matter (red hatched).
The red line denotes the upper limit (\ref{stablelimitG}), the magenta line shows the 
$\Lambda_{\rm mess}$-limit (\ref{Llimit1}) with $g=1$, and the blue lines represent the lower bound 
(\ref{phieqboundg}) for each value of $k$ shown. The dark matter $Q$ ball is excluded by the 
BAKSAN experiment (dark green hatched) \cite{Arafune} and IceCube (light green hatched) \cite{IC}.
Thermally produced gravitino dark matter is shown in brown lines for each value of $k$, and the 
dashed line corresponds to $m_{3/2}<100$~keV.
\label{fig1}}
\end{figure}

Next we consider the new type of $Q$ ball. Again, the $Q$ ball should be (a) stable to be the dark matter, 
and (b) have the correct DM abundance. In this case, since the potential must be dominated by 
$V_{\rm grav}$ at the onset of the field oscillation, we also have (c) $\phi_{\rm osc} > \phi_{\rm eq}$. 

The $Q$-ball abundance is estimated as
\begin{equation}
\frac{\rho_Q}{s} = \frac{3T_{\rm RH}}{4} \left.\frac{\rho_Q}{\rho_{\rm inf}}\right|_{\rm osc}
\simeq \frac{9}{4} T_{\rm RH} \left(\frac{\phi_{\rm osc}}{M_{\rm P}}\right)^2,
\end{equation}
where we use $\rho_Q = M_Q n_Q = m_{3/2} m_{\rm eff} \phi_{\rm osc}^2 = m_{3/2}^2 \phi_{\rm osc}^2$ and 
$\rho_{\rm inf}=3H_{\rm osc}^2M_{\rm P}^2= m_{3/2}^2 M_{\rm P}^2/3$ at the onset of the field oscillation.
Using $\rho_{\rm DM}/s=4.4\times 10^{-10}$~GeV, we need the field amplitude when the oscillation starts to be
\begin{equation}
\phi_{\rm osc} = 4.75 \times 10^{13} \, {\rm GeV} \left(\frac{T_{\rm RH}}{\rm GeV}\right)^{-1/2}.
\label{phioscDMn}
\end{equation}
Together with the condition (c) $\phi_{\rm osc} > \phi_{\rm eq}$, the reheating temperature has an
upper limit as
\begin{equation}
T_{\rm RH} < 1.13 \times 10^3 \, {\rm GeV} 
\left(\frac{M_F}{10^6\, {\rm GeV}}\right)^{-4}\left(\frac{m_{3/2}}{\rm GeV}\right)^2.
\label{rhoQphieqn}
\end{equation}
We then obtain the $k$-dependent upper limit on $T_{\rm RH}$ by inserting Eq.(\ref{m32MF}) into 
Eq.(\ref{rhoQphieqn}): 
\begin{equation}
T_{\rm RH} < 1.63 \times 10^{-6} \, {\rm GeV} g^{-2} k^{-2}.
\label{kdepTRH}
\end{equation}
The stability condition (a) is given by
\begin{equation}
m_{3/2} < b m_N = 0.333 \, {\rm GeV} \left(\frac{b}{1/3}\right),
\label{m32st}
\end{equation}
where Eq.(\ref{omegaQn}) is used.

The $Q$ ball can be the dark matter in the rectangular region surrounded by the stability 
condition (\ref{m32st}) in the red line, the lower limit on $m_{3/2}$ Eq.(\ref{Llimit2}) in magenta lines, 
the condition (c) Eq.(\ref{kdepTRH}) in blue lines, and $T_{\rm RH} \gtrsim$ 1~MeV, as shown in Fig.2. 
Varying the value of $k$, the intersection of Eqs.(\ref{Llimit2}) and (\ref{kdepTRH}) spans on the line
\begin{equation}
T_{\rm RH} < 4.42 \times 10^8 \, {\rm GeV} \left(\frac{m_{3/2}}{\rm GeV}\right)^2,
\label{TRHphieqn}
\end{equation}
shown in a black line in the figure. We thus hatched the triangle surrounded by Eqs.(\ref{TRHphieqn}) and 
(\ref{m32st}) and $T_{\rm RH} \gtrsim$ 1~MeV as the allowed region.

\begin{figure}[ht!]
\hspace{-10mm}
\includegraphics[width=90mm,angle=90]{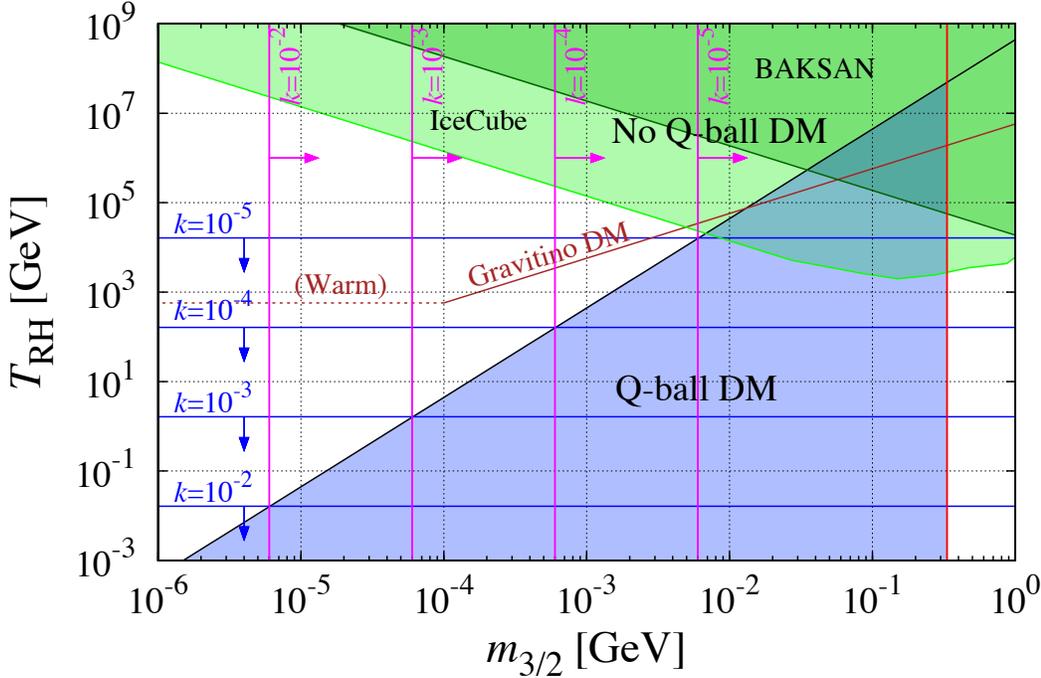} 
\vspace{10mm}
\caption{Allowed region for the new type $Q$ ball as the dark matter (blue hatched).
The red line denotes the stability condition (\ref{m32st}), the magenta lines show the 
$\Lambda_{\rm mess}$-limit (\ref{Llimit2}) with $g=1$ for each $k$, and the blue lines represent 
the upper bound (\ref{kdepTRH}) for each value of $k$. The trace of the intersection of Eqs.(\ref{Llimit2})
and (\ref{kdepTRH}), which is Eq.(\ref{TRHphieqn}), is shown in a black line.
The dark matter $Q$ ball is excluded by the 
BAKSAN experiment (dark green hatched) \cite{Arafune} and IceCube (light green hatched) \cite{IC}.
Thermally produced gravitino dark matter is shown in a brown line, and the 
dashed line corresponds to $m_{3/2}<100$~keV.
\label{fig2}}
\end{figure}

\subsection{Direct detection of dark matter $Q$ balls}
$Q$ balls can be detected through the so-called KKST (Kusenko-Kuzmin-Shaposhnikov-Tinyakov) process
\cite{KKST, Arafune, KK3}, which is similar to the Rubakov-Callan effect for the magnetic monopole \cite{RC}. 
When nucleons collide with a $Q$ ball, they enter the surface layer of the $Q$ ball,
and dissociate into quarks, which are converted to squarks. During the course of the process, 
the $Q$ ball releases energy of $\sim 1$~GeV per collision by emitting soft pions. Therefore, charged
particles are created along the path of the $Q$ ball through the detector. Since the $Q$ balls in our
dark halo have the velocity $v \sim 10^{-3}$, those experiments for the subrelativistic monopole search 
can be applied to $Q$-ball detection.

The observable is the upper limit of the $Q$-ball flux at the certain cross section, since no $Q$ ball
has been detected so far. The flux and the cross section of the dark matter $Q$ ball are given respectively by
\begin{equation}
F < F_{\rm DM} \simeq \frac{\rho_{\rm DM} v}{4\pi M_Q}, \qquad
\sigma_Q \simeq \pi R_Q^2,
\end{equation}
where $\rho_{\rm DM} \sim 0.3$~GeV/cm$^3$ is the local dark matter density, and 
they are related to the charge $Q$ though the $Q$-ball properties Eqs.(\ref{MQg}) $-$ (\ref{omegaQg})
for the gauge-mediation type and Eqs.(\ref{MQn}) $-$ (\ref{omegaQn}) for the new type. 
In Fig.3, we show the theoretically expected regions for the gauge-mediation and new types of $Q$ ball
respectively by red and blue triangle areas. Also shown are the upper limits of the $Q$-ball flux from 
BAKSAN \cite{Arafune} and IceCube \cite{IC},\footnote{
We simply extrapolate the flux at the cross section $\sim 10^{-22}$~cm$^{-2}$sec$^{-1}$sr$^{-1}$ to larger
cross sections.}
hatched in dark green and green, respectively. 
We see that some region for the new type of $Q$ ball is already excluded and there is a good
possibility of direct detection of the $Q$ ball in the near future.

\begin{figure}[ht!]
\hspace{-10mm}
\includegraphics[width=90mm,angle=90]{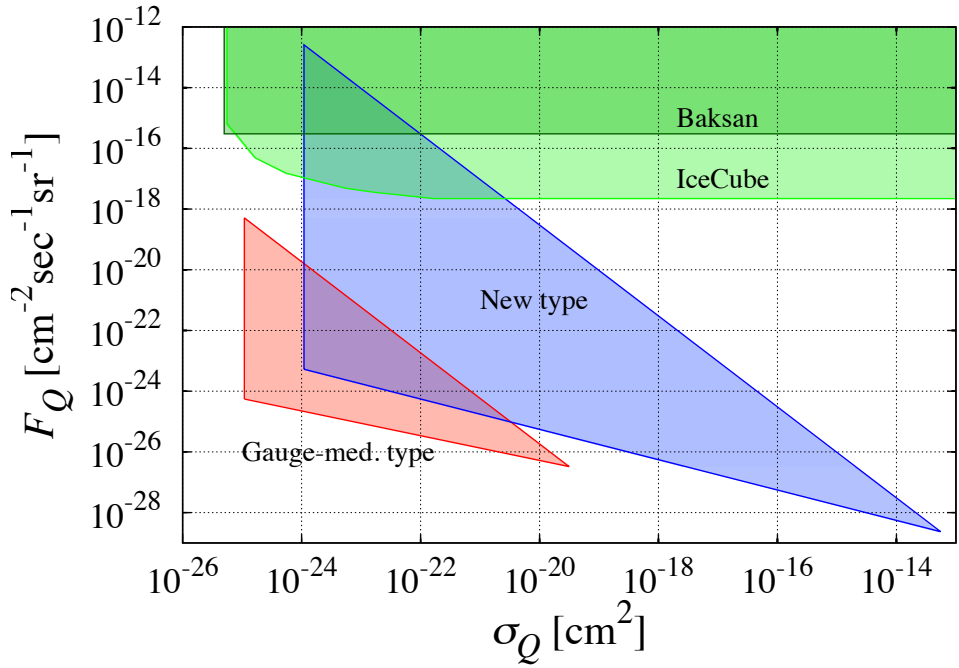}
\vspace{10mm}
\caption{Expected flux of the dark matter $Q$ ball for the gauge-mediation (red) and the new (blue) types.
We also plot the flux limits from the BAKSAN (dark-green) \cite{Arafune} and IceCube (green) \cite{IC} 
experiments.
\label{fig3}}
\end{figure}

Finally, we use these experimental bounds to impose some constraints on the reheating temperature 
by using Eqs.(\ref{Qg}) and (\ref{phioscDMg}) for the gauge-mediation type in Fig.1, while using 
Eqs.(\ref{Qn}) and (\ref{phioscDMn}) for the new type in Fig.2.
In this way, we plot the regions where the dark matter $Q$ ball is excluded by the 
BAKSAN experiment (dark green) \cite{Arafune} and IceCube (light green) \cite{IC}.
For the gauge-mediation type $Q$ ball in Fig.1, the excluded regions are getting closer to
the theoretically expected area. On the other hand, for the new type $Q$ balls in Fig.2, 
the upper part of the allowed region is now observationally excluded, and $Q$-ball dark matter
can be realized for $T_{\rm RH} \lesssim 10^4$~GeV.

\subsection{Comparing with gravitino dark matter}
As mentioned above, the gravitino is also a natural candidate for the dark matter in gauge mediation.
We therefore estimate the parameters for which gravitinos account for the dark matter and 
show that the region is totally different from that where the $Q$ ball is the dominant component 
of the dark matter. The abundance of gravitinos thermally produced by scatterings is given by \cite{KKMY}
\begin{equation}
Y_{3/2}= 7.67\times 10^{-9}  \left(\frac{T_{\rm RH}}{10^8\, {\rm GeV}}\right) 
\left(\frac{m_{\tilde{g}}}{\rm TeV}\right)^2\left(\frac{m_{3/2}}{\rm GeV}\right)^{-2},
\end{equation}
for $m_{3/2}\ll m_{\tilde{g}}$, where $m_{\tilde{g}}$ is the gaugino mass. Since 
$\rho_{\rm DM}/s=4.4\times 10^{-10}$~GeV and $\rho_{3/2} \le \rho_{\rm DM}$, we have
\begin{equation}
T_{\rm RH} \le 5.74\times 10^6 \, {\rm GeV} 
\left(\frac{m_{\tilde{g}}}{\rm TeV}\right)^{-2}\left(\frac{m_{3/2}}{\rm GeV}\right) 
\quad (100\, {\rm keV} \lesssim m_{3/2} \lesssim {\rm GeV}).
\label{rho32TRH}
\end{equation}
This is shown in a brown line in Fig.2. If the gravitino mass is less than 100 keV,
the gravitino is mainly produced by the decay of the SUSY particle \cite{MMY}. Then the upper limit on
the reheating temperature becomes of the order of the mass of SUSY particles for 
keV $\lesssim m_{3/2} \lesssim$ 100 keV, although the gravitino dark matter may be too warm. 
Therefore, the new-type $Q$ ball can be the dark matter for a lower reheating temperature than the 
gravitino dark matter case, and even for the lighter gravitino mass region such as 
keV $\lesssim m_{3/2} \lesssim$ 100 keV.

On the other hand, comparing to the gauge-mediation type $Q$ ball, we need to rephrase Eq.(\ref{rho32TRH})
in terms of $M_F$. Using Eq.(\ref{m32MF}), we obtain the constraint as
\begin{equation}
T_{\rm RH} \le 2.18 \times 10^2 \, {\rm GeV} g^{-1} k^{-1}
\left(\frac{m_{\tilde{g}}}{\rm TeV}\right)^{-2}\left(\frac{M_F}{10^6\, {\rm GeV}}\right)^2 
\quad (1.62\times 10^6 \, {\rm GeV} k^{-1} \lesssim M_F),
\label{32TRH}
\end{equation}
which is shown in brown lines for each $k$ in Fig.1. As before, the constraint on $T_{\rm RH}$ stays
almost constant and of the order of the SUSY particle mass for $m_{3/2} \lesssim$ 100 keV, as displayed by
the brown dashed line in the figure. As can be seen, this type of $Q$ ball can be 
the dark matter for $T_{\rm RH} \lesssim 10^3$~GeV, whereas the gravitino can be the (cold) dark matter
for the opposite case $T_{\rm RH} \gtrsim 10^3$~GeV. Also the $Q$-ball dark matter can exist for 
60 eV $\lesssim m_{3/2} \lesssim$ keV, in particular, for $k\simeq 1$. See Eq.(\ref{m32MF}).

\section{Conclusion}
We have investigated $Q$-ball dark matter in gauge-mediated SUSY breaking, and sought 
its direct detection. We have found that the $Q$ ball can be the dark matter in the region of parameter 
space different from that for the gravitino dark matter. The $Q$ ball will be the dark matter for the lower 
reheating case such as $T_{\rm RH} \lesssim 10^4$~GeV, whereas a higher $T_{\rm RH}$ is necessary 
for gravitino dark matter. Moreover, dark matter $Q$ balls could, in the future, be detected directly in the 
large-volume neutrino telescopes, such as IceCube \cite{IC}, Baikal \cite{Baikal}, KM3NeT \cite{KM3NeT}, 
and so on. In particular, the IceCube experiment may detect the $Q$ ball with their full volume 
and the slow-particle trigger in the near future \cite{IC}. This is a distinctive feature of $Q$-ball dark matter,
whereas gravitino dark matter cannot be found in collider experiments.

Finally, we make a comment on baryogenesis. The $Q$-ball dark matter scenario
requires a low reheating temperature ($T_{\rm RH} \lesssim 10^4$~GeV). 
Affleck-Dine baryogenesis usually works at such a low reheating temperature.
In this case, another flat direction than that which forms dark matter $Q$ balls will produce the baryon 
asymmetry of the universe \cite{KK5}. 

Alternatively, nonthermal leptogenesis by right-handed neutrinos with nearly degenerate masses 
may operate at such low reheating temperature, although it should be higher than $\sim 10^2$~GeV. 
Affleck-Dine leptogenesis \cite{MuYa} may also work for $T_{\rm RH} \gtrsim 10^2$~GeV. 
In these cases, IceCube may detect $Q$-ball dark matter in very near future experiments 
(see Figs.1 and 2).

\section*{Acknowledgments}
This work is supported by Grant-in-Aid for Scientific Research  
23740206 (S.K.), 25400248 (M.K.), 26104009 (T.T.Y) and 26287039 (T.T.Y.) 
from the Ministry of Education, Culture, Sports, Science, and 
Technology (MEXT), Japan, and also by the World Premier International Research 
Center Initiative (WPI Initiative), MEXT, Japan (M.K. and T.T.Y.).



\end{document}